\documentclass{acta}

\usepackage{amsmath}
\usepackage{amssymb}
\usepackage{rotating}
\usepackage[hidelinks]{hyperref} 
\usepackage{bm}

\usepackage{xcolor}

\newcommand{\ra}[4]{${#1}^{\rm h}{#2}^{\rm m}{#3}\fs{#4}$}
\newcommand{\dec}[4]{${#1}\arcdeg{#2}\arcmin{#3}\farcs{#4}$}

\newcommand\fs{\mbox{$.\!\!^{\mathrm s}$}}%
\newcommand\arcdeg{\mbox{$^\circ$}}%
\newcommand\arcmin{\mbox{$^\prime$}}%
\newcommand\farcs{\mbox{$.\!\!^{\prime\prime}$}}%

\newcommand{\dummy}[1]{}

\begin{document}

\begin{Titlepage}

\Title{\textit{Eppur non si trovano Vol.~2}:\\
No Planetary-mass Primordial Black Holes\\ 
toward the Andromeda Galaxy}


\Author{Przemek Mr\'oz$^1$ and Andrzej Udalski$^1$}
{$^1$Astronomical Observatory, University of Warsaw, Al. Ujazdowskie 4, 00-478 Warszawa, Poland\\
e-mail: pmroz@astrouw.edu.pl, udalski@astrouw.edu.pl}

\Received{MM DD, YYYY}

\end{Titlepage}

\vspace*{-10pt}
\Abstract{
A recent preprint by Sugiyama {\it et al.}~reported the discovery of twelve candidates for short-timescale (less than one day) gravitational microlensing events based on high-cadence photometric observations of the Andromeda Galaxy (M31) using the Subaru Hyper Suprime-Cam. These detections were attributed to a large population of planetary-mass primordial black holes (PBHs) that could account for the entirety of the dark matter in the Milky Way and M31 halos. However, these results are in clear tension with previous searches for short-timescale microlensing events toward the Magellanic Clouds, such as those by the OGLE survey. In addition, both the temporal and spatial distributions of the Subaru candidates are inconsistent with expectations for microlensing events. 

Here, we reanalyze the Subaru data using an independent difference image analysis photometric pipeline. We find that all twelve candidates identified by Sugiyama {\it et al.}~exhibit asymmetric light curves and/or variability on multiple nights of Subaru observations. Our analysis reveals that among them ten objects are RR Lyrae stars, one is an eclipsing binary, and one is an unclassified variable star. 

We find no compelling evidence for short-timescale microlensing events among the candidates identified in the Subaru data set, nor for a significant population of planetary-mass PBHs as dark matter components. Our results underscore the necessity of robust variable-star rejection in high-cadence microlensing searches using large telescopes. 
}
{Gravitational microlensing (672), Dark matter (353), Milky Way dark matter halo (1049), Primordial black holes (1292), RR Lyrae variable stars (1410), Andromeda Galaxy (39)}

\section{Introduction} \label{sec:intro}

A recent preprint by \citet{sugiyama2026} reported the discovery of twelve candidates for short-timescale gravitational microlensing events (among them, four were classified as ``secure'' candidates) based on an analysis of high-cadence photometric observations of the Andromeda Galaxy (M31) carried out with the Subaru Hyper Suprime-Cam (HSC). These detections, if real, cannot be easily explained by gravitational microlensing events from known planetary, substellar, and stellar populations. Instead, \citet{sugiyama2026} proposed that these candidates represent a large population of planetary-mass primordial black holes (PBHs) which could account for a substantial fraction of dark matter in the halos of both the Milky Way and M31. According to their analysis, four ``secure'' detections would imply that PBHs with masses ranging from $10^{-7}$ to $10^{-6}$\,M$_{\odot}$ constitute 6--40\% of dark matter, while all twelve candidates would support the hypothesis that planetary-mass PBHs make up 25--80\% of dark matter.

However, these results contradict earlier studies that focused on searching for short-timescale gravitational microlensing events toward the Milky Way dark matter halo. \citet{renault1997,renault1998} analyzed high-cadence observations of the Magellanic Clouds carried out as part of the EROS survey. They found that PBHs with masses between $10^{-7}$ and $10^{-6}$\,M$_{\odot}$ cannot make up more than 40--10\% of the Milky Way dark matter halo, respectively. Similar conclusions were reached by \citet{alcock1996,alcock1998}.

These limits were substantially improved in a recent study by \citet{mroz2024d}, who analyzed high-cadence observations of nearly 35 million stars in the Magellanic Clouds, collected from 2022 October through 2024 May as part of the Optical Gravitational Lensing Experiment (OGLE; \citealt{udalski2015}). No short-timescale events were detected in the OGLE data, indicating that PBHs and other compact objects with masses ranging from $1.4\times 10^{-8}$ to 0.013\,M$_{\odot}$ may comprise at most 1\% of dark matter. (All these limits are 95\% confidence level upper limits and were derived under the assumption of a monochromatic PBH mass function.) \citet{sugiyama2026} and \citet{mroz2024d} results are in a clear tension: had planetary-mass PBHs constituted the entirety of dark matter, OGLE would have discovered over a thousand short-timescale microlensing events in its high-cadence observations of the Magellanic Clouds. However, no such events were observed.

Additionally, the findings reported by \citet{sugiyama2026} contradict an earlier study that used a subset of the Subaru M31 observations: \citet{niikura2019} reported the detection of only one microlensing event candidate and derived very strong limits on the abundance of planetary-mass PBHs in dark matter.

Moreover, some general properties of the objects reported by \citet{sugiyama2026} are inconsistent with the characteristics of microlensing events. Although the Subaru observations spanned ten nights across 2014, 2017, and 2020 (and useful photometry could be extracted for only eight of these nights), \citet{sugiyama2026} reported the discovery of twelve candidate microlensing events that occurred during just two nights of 2014-11-24 and 2017-09-20 (with a total effective observation time of 13.5 hours). However, no candidate events were identified in the additional data collected in 2020, which included a total of 25.8 hours of observations. Assuming that events occur uniformly over time, the expected number of events during the 2020 observing run is $12 \times 25.8 / 13.5 = 22.9$, if the event detection efficiency in 2020 is similar to that in the 2014 and 2017 data. The Poisson probability of observing zero events when 22.9 are expected is $p=e^{-22.9} = 1.1 \times 10^{-10}$.

The temporal and spatial distributions of the candidates are equally problematic. Given the cadence and sampling of Subaru observations, we expect that the peak times of genuine microlensing events should be uniformly distributed over time. However, all five candidate events that occurred during the night of 2014-11-24 peaked within 17\,minutes of each other. Similarly, among the seven candidate events that took place during the night of 2017-09-20, six peaked within 32\,minutes of one another. In addition, in nature, the distribution of the normalized impact parameter $z_0 \equiv u_0/\rho$ should be exactly uniform, where $u_0$ and $\rho$ are the impact parameter and source star radius in Einstein radius units. However, 75\% of candidate events reported by \citet{sugiyama2026} have $z_0 > 0.75$, indicating a significant deficit of events with small $z_0$.

We also expect that the spatial distribution of microlensing events in the sky should reflect the distribution of source stars in M31, with the majority of genuine events occurring in the M31 disk. That is due to two factors. First, thanks to the favorable geometry of M31, the microlensing optical depth should be greater when observing the far side of the M31 disk, because the sightline passes through the denser inner halo of M31 \citep{crotts1992,jetzer1994}. Second, the number of potential source stars is larger in the disk. However, all twelve microlensing candidates identified by \citet{sugiyama2026} are instead located in the M31 halo, far from its disk.

Despite these issues,  some studies have accepted the \citet{sugiyama2026} results as evidence that planetary-mass PBHs constitute the bulk of dark matter. 
They set out to investigate the origins of this alleged PBH population. \citet{domenech2026} suggested that a broad enhancement of the curvature power spectrum could simultaneously explain the formation of planetary-mass PBHs and the nanohertz stochastic gravitational-wave background reported by pulsar timing arrays. Meanwhile, \citet{kasai2026} proposed that this purported population of planetary-mass PBHs could arise from an enhancement of curvature perturbations on small scales in the axionlike curvaton models, where the curvaton is a field responsible for generating primordial curvature perturbations instead of the inflaton. \citet{blas2026} discussed the possibility of detecting gravitational wave background in the $\mu$Hz frequency range generated by the collapse of primordial curvature perturbations into planetary-mass PBHs with the proposed lunar detectors. 

\citet{sugiyama2026} presented only partial light curves of their microlensing event candidates, which covered just one night of observations. A close examination of all twelve light curves (shown in Figure~10 of \citealt{sugiyama2026}) reveals that many of them are asymmetric, featuring a rapid rise followed by a slower decline. This contrasts with the expected symmetric profile of microlensing events and raises doubts about whether the detected objects are true microlensing events or rather more mundane variable stars. Considering the other unusual characteristics of these purported events and the potential importance of the detection of short-timescale microlensing events in the direction of M31, we decided to extract full light curves using an independent photometric pipeline. This article presents the results of our findings.

\section{Data}

We used exactly the same data set as analyzed by \citet{sugiyama2026}. The M31 Galaxy was observed by Subaru HSC \citep{miyazaki2018,komiyama2018,kawanomoto2018,furusawa2018} on ten nights in 2014, 2017, and 2020. However, useful photometry could be extracted for only eight nights by \citet{sugiyama2026}, and we used data from exactly the same nights. The log of Subaru HSC observations is presented in Table~\ref{tab:log}. Thanks to the HSC field of view of $1.5\,\mathrm{deg}^2$, almost the entire M31 disk can be captured with just one pointing. The pixel size of the camera is $0\zdot\arcs169$ (at the field center).

We retrieved the raw Subaru HSC data and the corresponding calibration images (bias, dark, and flat frames) from the Subaru Mitaka Okayama Kiso Archive\footnote{https://smoka.nao.ac.jp/} (SMOKA; \citealt{baba2002}). We used standard procedures from the \textsc{ccdproc} package \citep{craig2025} to subtract bias and dark frames and perform the flat-field correction; data from each night were processed separately.

All images were taken with an exposure time of 90\,s, which, in combination with a readout time of 30\,s, results in a 2-minute cadence. The number of analyzed images collected during a given night ranges from 44 to 214, and it may be slightly lower than the values shown in Table~\textsc{I} of \citet{sugiyama2026} because some frames were discarded due to poor quality.
All images collected on the night of 2014-11-24 were taken through the $r$-band filter, while the remaining images were taken through the $r2$-band filter. Although both the $r$ and $r2$ filters have similar transmission curves, the $r2$ filter offers improved uniformity \citep{kawanomoto2018}. The median seeing during each night (as measured on the HSC detector 039) varies from $0\zdot\arcs69$ (2014-11-24) to $1\zdot\arcs14$ (2020-11-12). Table~\ref{tab:log} provides information on the number of analyzed images and median seeing on each night of observations.

\begin{table}
\caption{Log of Subaru HSC Observations}
\label{tab:log}
\centering
\begin{tabular}{cccc}
\hline \hline
Date & Filter & Images & FWHM\\
\hline
2014-11-24 & $r$  & 186 & $0\zdot\arcs69$\\
2017-09-20 & $r2$ & 214 & $0\zdot\arcs78$\\
2020-10-21 & $r2$ & 129 & $0\zdot\arcs76$\\
2020-10-22 & $r2$ & 141 & $0\zdot\arcs78$\\
2020-11-11 & $r2$ & 44  & $0\zdot\arcs91$\\
2020-11-12 & $r2$ & 90  & $1\zdot\arcs14$\\
2020-11-14 & $r2$ & 152 & $0\zdot\arcs83$\\
2020-11-20 & $r2$ & 134 & $1\zdot\arcs00$\\
\hline
\end{tabular}
\end{table}


The Subaru HSC images of M31 resemble, to some extent, the images collected by the OGLE survey, particularly in terms of PSF sampling and stellar crowding in the observed fields. Thus, we decided to re-reduce the Subaru images of the microlensing event candidates using our well-tested OGLE photometry pipeline \citep{udalski2003}, which provides very precise photometry even in the densest stellar fields. This required tuning the pipeline parameters to account for the differences between the OGLE and Subaru images. The pipeline is based on the version of the Difference Image Analysis (DIA) algorithm implemented by \citet{wozniak2000}.

The CCDs of the Subaru HSC camera have $2048\times4176$ pixels
 and are naturally divided into four subsections. Each subsection is read by
one of the four amplifiers, resulting in an output of each
amplifier having $512\times4176$ pixels. We treat each of these subsections as a separate entity in our reductions. The corresponding field covered by each subsection is named M31\_A.R.CC, where A is the amplifier number (from 1 to 4), R indicates the filter, and CC is the detector number (00-99; we excluded four severely vignetted edge CCDs) according to the Subaru HSC schematic and nomenclature (see Figure~1 of \citealt{aihara2018}). Thus, each CCD detector from the Subaru HSC covers four of our fields.

DIA photometric reductions for each field were carried out on two   
subframes, each measuring $512\times2088$ pixels. The reference image
for each field was constructed from 15 individual images characterized by the best 
seeing, highest image quality, and lowest background. Exceptionally favorable seeing conditions occurred during a few hours on the night of 2014-11-24. Several images were taken with seeing
better than 0\zdot\arcs5. The images collected during this period were used
to construct the reference images. Thus, the reference image epoch
is ${\rm HJD'} = {\rm HJD}-2\,500\,000 = 6985.82$.

Individual images were aligned, resampled, and rescaled to match the
photometric level of the first image of the stack. They were then co-added,
creating a nearly noiseless, very deep reference image of each of  
the analyzed fields. Although the Subaru images are occasionally
contaminated by long, faint traces of cosmic-ray hits, we chose not to
remove them because our algorithms (which perform well on the OGLE images) did not
work correctly on this specific type of cosmic ray trace and could introduce  
some artifacts. The first image from the stack defines the $(x,y)$ coordinate
system for subsequent reductions and sets the instrumental photometry level of  
the reference image. The reference image objects, their coordinates, and
reference flux levels were then extracted using the PSF photometry program \textsc{DoPhot} \citep{schechter1993}.

We used the same reference image for the reductions of all frames in a given field. Although we were aware that the {\it r} filter used on 
the reference image night of 2014-11-24 was later replaced by the
{\it r2} filter for the subsequent nights, the characteristics of the two filters are so similar, as already mentioned,
that the replaced filter should not introduce any significant error. Our
choice is supported by our experience with joint reductions of
\mbox{OGLE-III} and \mbox{OGLE-IV} images, which were taken through filters with considerably larger differences than in this case \citep{mroz2024a}.

Each individual image was processed as follows. First, after
finding the transformation between the reference and local coordinate grids,
the image was resampled to match the reference image coordinate system. Then, the transformation between the reference and target images was derived bringing them to the same seeing and photometric level, and they were subtracted to produce the difference image. The resulting difference images turned out to be of excellent quality across the entire subframe, and even the dense
stellar regions of M31 were subtracted smoothly.

Photometry of each star from the reference image was then derived on the
difference image with the PSF photometry at the reference position
of each star. We used a slightly smaller radius for the PSF fitting (3~pixels) than in the standard OGLE reductions, as this yielded photometry with a slightly better quality in the dense and faint stellar M31 fields. The difference flux of each star was then added to its
reference flux, converted to the magnitude system, and rescaled to the
instrumental system, roughly resembling the standard $r$-band system by
adding a constant -- the magnitude scale zero point. The instrumental
photometry was stored in the databases similar to those used in the OGLE
project for easy data manipulation and photometric exploration of M31. 

To maintain high quality of photometry, we limited our sample of Subaru 
HSC images to those with seeing  
smaller than 1\zdot\arcs3 (images with poorer seeing may cause the pipeline to fail due to an insufficient number of transformation stars; in addition, the photometric
errors rapidly increase with increasing seeing). We also removed several almost empty images that were
evidently taken through thick clouds. This filtering reduced our sample 
by about 10\%, leaving 1090 images.


We noticed that the World Coordinate System transformations supplied with the raw FITS files from the SMOKA archive were only approximate. Thus, we cross-matched the positions of the brightest stars with the \textit{Gaia}~DR3 catalog \citep{gaia2016,gaia_edr3}. This step allowed us to derive accurate quadratic polynomial transformations between the \textit{Gaia} equatorial coordinates (in the \textit{Gaia} celestial reference frame) and the reference image coordinates. The typical rms scatter of the residuals from these transformations was better than 0.2--0.3 HSC pixels. We then used these transformations to determine the positions of the microlensing candidates on our reference images based on their equatorial coordinates, as provided in Table~\textsc{VII} of \citet{sugiyama2026}.

Finally, using the coordinate transformations, we cross-matched stars in the reference images with the Pan-STARRS1 (PS1) DR2 catalog \citep{chambers2016} to calibrate the magnitudes. We used magnitudes from the PS1 Stack Object catalog, as we found that the calibration uncertainties are smaller than when using the PS1 Mean catalog. The typical zero-point accuracy varies between 0.01 and 0.03 mag, depending on the specific field.

\section{Niikura \textit{et al.}~(2019) Candidate Event}
\label{sec:niikura}

\citet{niikura2019} presented the results of the searches for microlensing events using only one night of high-cadence M31 observations (from the night of 2014-11-24), which is a small subset of images analyzed by \citet{sugiyama2026}. They presented the discovery of only one microlensing event candidate, located at the equatorial coordinates of (RA, Decl.) = (\ra{00}{45}{33}{413}, \dec{+41}{07}{53}{03}), for the J2000 epoch. With only one night of Subaru observations, they were unable to rule out that this was just a mundane variable star. Interestingly, this purported short-timescale event is not included in the sample of event candidates found by \citet{sugiyama2026} (although the latter authors detected five candidate events on the night of 2014-11-24 that have been missed or discarded by \citealt{niikura2019}).

\begin{figure}[h]
\includegraphics[width=\textwidth]{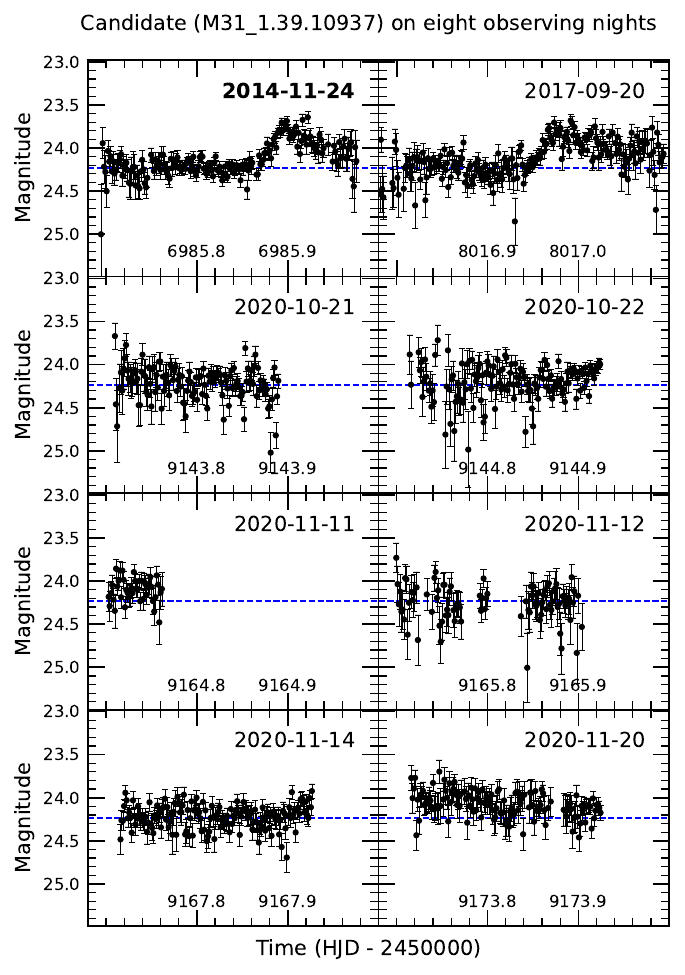}
\caption{Full Subaru HSC light curve of the candidate microlensing event identified by \citet{niikura2019} reveals significant variability on three nights (2014-11-24, 2017-09-20, and 2020-11-20), consistent with an RR Lyrae star rather than a microlensing event.}
\label{fig:lens99}
\end{figure}

Figure~\ref{fig:lens99} shows our full Subaru HSC light curve of that star, which exhibits significant brightness variations during at least three nights: 2014-11-24, 2017-09-20, and 2020-11-20. The top left panel can be directly compared with Figure~4 of \citet{niikura2019}. It can be immediately seen that the photometry from our pipeline is of much better quality. 

\begin{figure}[h]
\includegraphics[width=\textwidth]{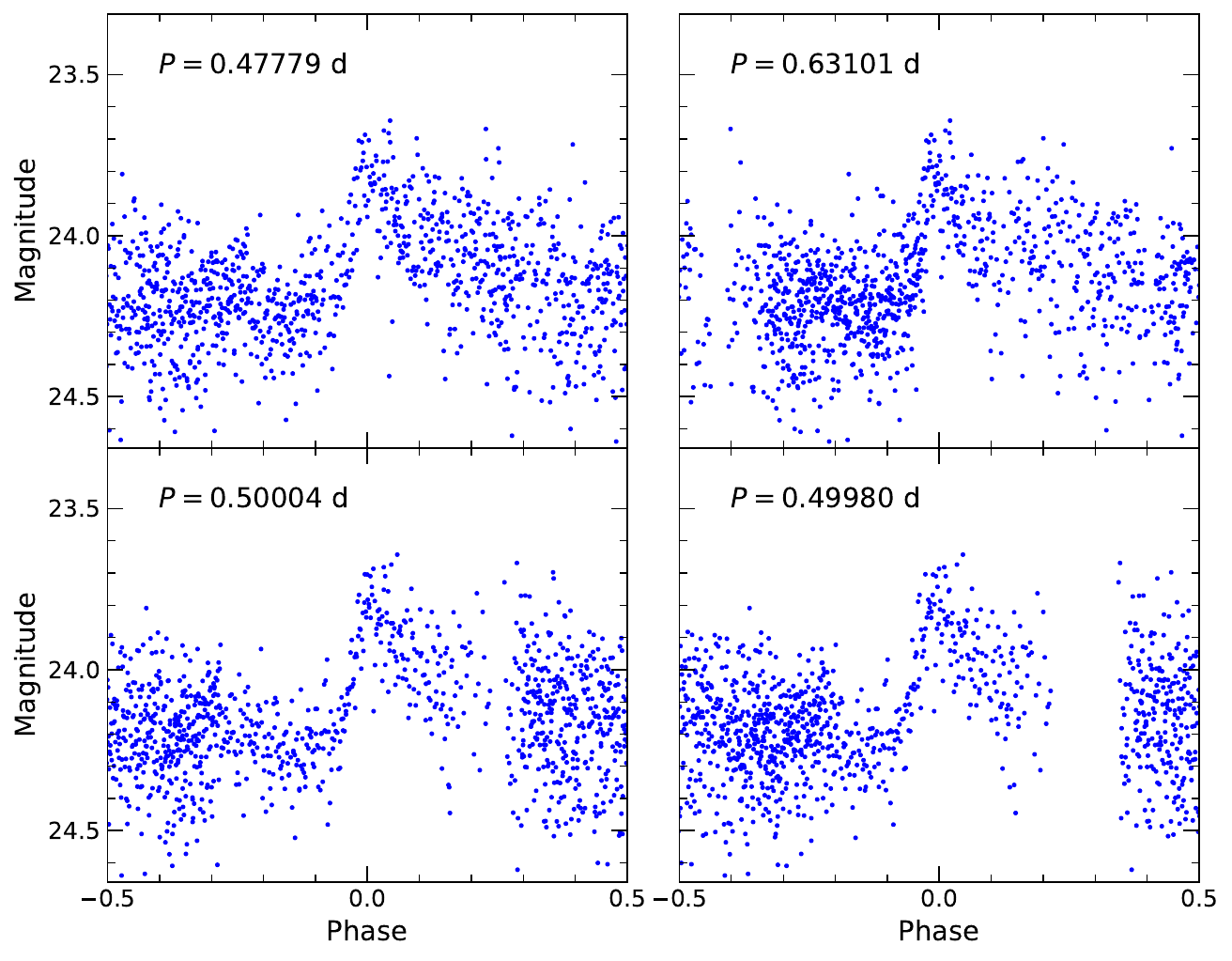}
\caption{Light curve of the candidate microlensing event identified by \citet{niikura2019} folded with four possible pulsation periods.}
\label{fig:phased_lens99}
\end{figure}

The variability of the microlensing candidate is consistent with that of an RR Lyrae star. Because of the relatively sparse light curve coverage, the data can be folded with many possible pulsation periods, including 0.47779, 0.63101, 0.50004, or 0.49980 days. The example phase-folded light curves corresponding to these periods are shown in Figure~\ref{fig:phased_lens99}.

\section{Sugiyama \textit{et al.}~(2026) Candidate Events}

\begin{figure}[h]
\centering
\includegraphics[width=.32\textwidth]{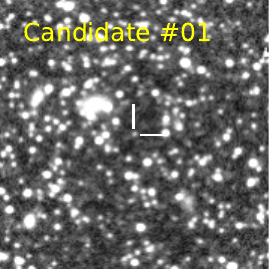}
\includegraphics[width=.32\textwidth]{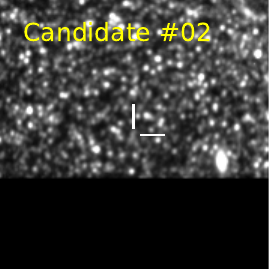}
\includegraphics[width=.32\textwidth]{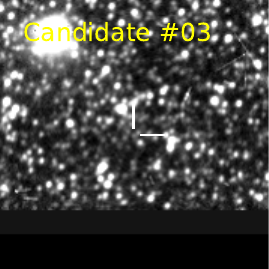}\\
\includegraphics[width=.32\textwidth]{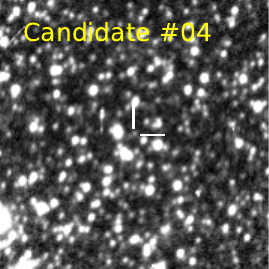}
\includegraphics[width=.32\textwidth]{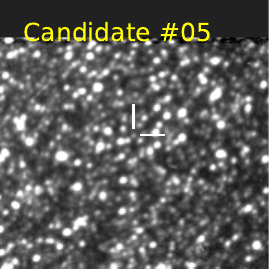}
\includegraphics[width=.32\textwidth]{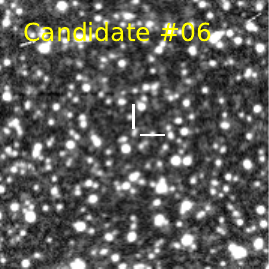}\\
\includegraphics[width=.32\textwidth]{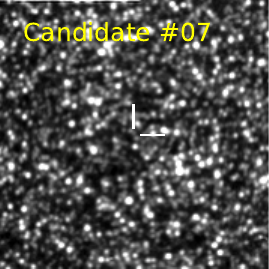}
\includegraphics[width=.32\textwidth]{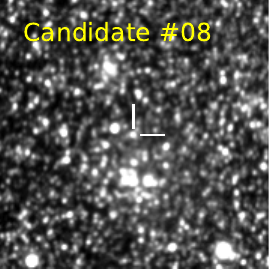}
\includegraphics[width=.32\textwidth]{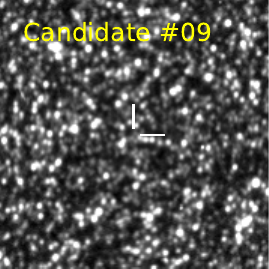}\\
\includegraphics[width=.32\textwidth]{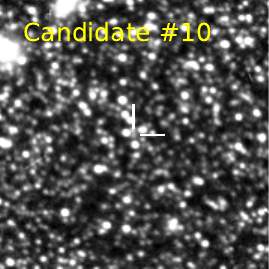}
\includegraphics[width=.32\textwidth]{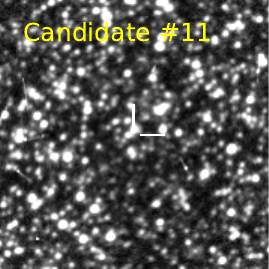}
\includegraphics[width=.32\textwidth]{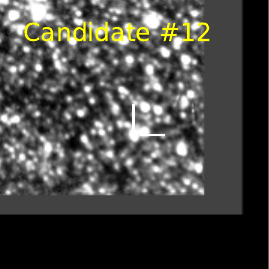}\\
\caption{Finding charts for the microlensing event candidates found by \citet{sugiyama2026}. Each image is $30'' \times 30''$, North is up and East is to the left.}
\label{fig:finding_charts}
\end{figure}

\citet{sugiyama2026} identified twelve short-timescale microlensing event candidates in the analyzed Subaru HSC data set. Their equatorial coordinates (taken from Table~\textsc{VII} of \citealt{sugiyama2026}) and discovery nights are reported in Table~\ref{tab:events}. Four candidates were flagged as ``secure'' by \citet{sugiyama2026} and they are marked with the symbol $\dagger$ in Table~\ref{tab:events}. The finding charts for all twelve candidates are presented in Figure~\ref{fig:finding_charts}. These images are $30''\times30''$ cutouts of the $r$-band reference images, North is up and East is to the left. The position of each candidate is indicated by the white tick marks.

\begin{figure}[h]
\centering
\includegraphics[width=.32\textwidth]{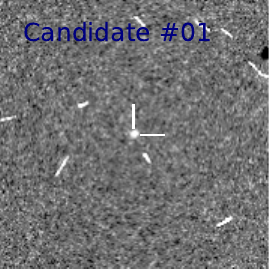}
\includegraphics[width=.32\textwidth]{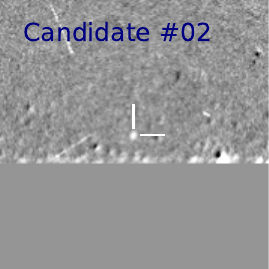}
\includegraphics[width=.32\textwidth]{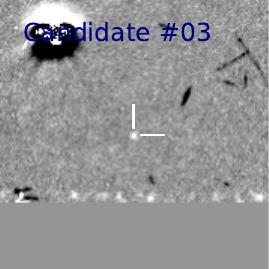}\\
\includegraphics[width=.32\textwidth]{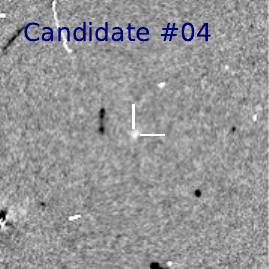}
\includegraphics[width=.32\textwidth]{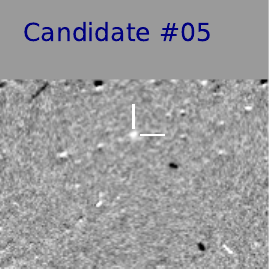}
\includegraphics[width=.32\textwidth]{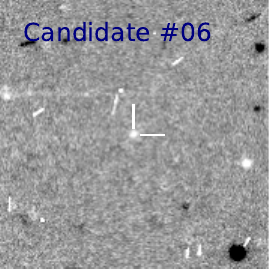}\\
\includegraphics[width=.32\textwidth]{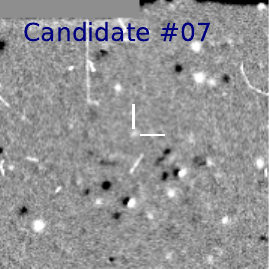}
\includegraphics[width=.32\textwidth]{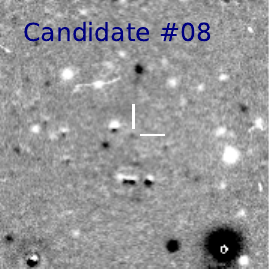}
\includegraphics[width=.32\textwidth]{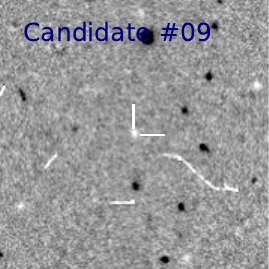}\\
\includegraphics[width=.32\textwidth]{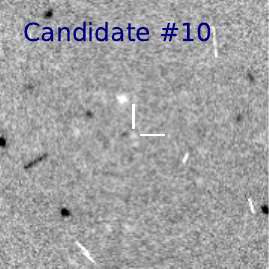}
\includegraphics[width=.32\textwidth]{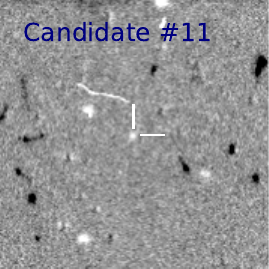}
\includegraphics[width=.32\textwidth]{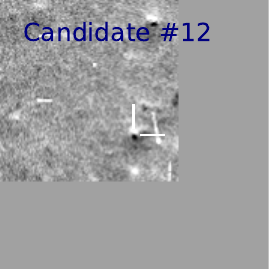}\\
\caption{Stacks of $\approx 15$ difference images centered on the peaks of the purported microlensing events. The position of each variable is marked by the white tick marks. Images are $30'' \times 30''$, North is up and East is to the left.}
\label{fig:diff_images}
\end{figure}

To confirm the identification of the targets, we created stacks of $\approx 15$ difference images taken around the peak of the purported microlensing events. In difference images, the light of all constant stars is subtracted, leaving only objects with changing brightness. This step allowed us to make a final verification of the identified stars. The $30'' \times 30''$ cutouts of the stacks of difference images are shown in Figure~\ref{fig:diff_images}.

\begin{table}
\caption{Candidate Microlensing Events Identified by \citet{sugiyama2026}}
\label{tab:events}
\centering
\begin{tabular}{ccccl}
\hline \hline
Event & Peak night & R.A. & Decl. & \multicolumn{1}{c}{Classification}\\
\hline
1$\dagger$  & 2014-11-24 & \ra{00}{46}{38}{41} & \dec{+41}{16}{43}{2} & RR Lyr (0.672922~d) \\
2$\dagger$  & 2014-11-24 & \ra{00}{42}{53}{02} & \dec{+40}{48}{07}{6} & RR Lyr (0.495750~d) \\
3           & 2014-11-24 & \ra{00}{41}{20}{91} & \dec{+41}{41}{16}{1} & RR Lyr (0.496301~d) \\
4           & 2014-11-24 & \ra{00}{41}{07}{94} & \dec{+41}{48}{54}{0} & RR Lyr (0.569439~d) \\
5           & 2014-11-24 & \ra{00}{40}{07}{90} & \dec{+41}{36}{17}{8} & RR Lyr (0.463168~d) \\
6           & 2017-09-20 & \ra{00}{46}{21}{70} & \dec{+41}{09}{17}{1} & RR Lyr (0.471390~d)\\
7           & 2017-09-20 & \ra{00}{44}{07}{74} & \dec{+41}{05}{17}{6} & RR Lyr (0.793001~d)\\
8           & 2017-09-20 & \ra{00}{42}{56}{85} & \dec{+40}{49}{09}{8} & Eclipsing star\\
9$\dagger$  & 2017-09-20 & \ra{00}{42}{09}{81} & \dec{+41}{44}{38}{3} & RR Lyr (0.535199~d)\\
10          & 2017-09-20 & \ra{00}{39}{47}{83} & \dec{+41}{11}{05}{0} & RR Lyr (0.493408~d)\\
11          & 2017-09-20 & \ra{00}{40}{13}{54} & \dec{+41}{31}{29}{5} & RR Lyr (0.547396~d)\\
12$\dagger$ & 2017-09-20 & \ra{00}{39}{33}{27} & \dec{+40}{50}{06}{6} & Variable star\\
\hline
\end{tabular}
\end{table}

\begin{figure}[h]
\includegraphics[width=\textwidth]{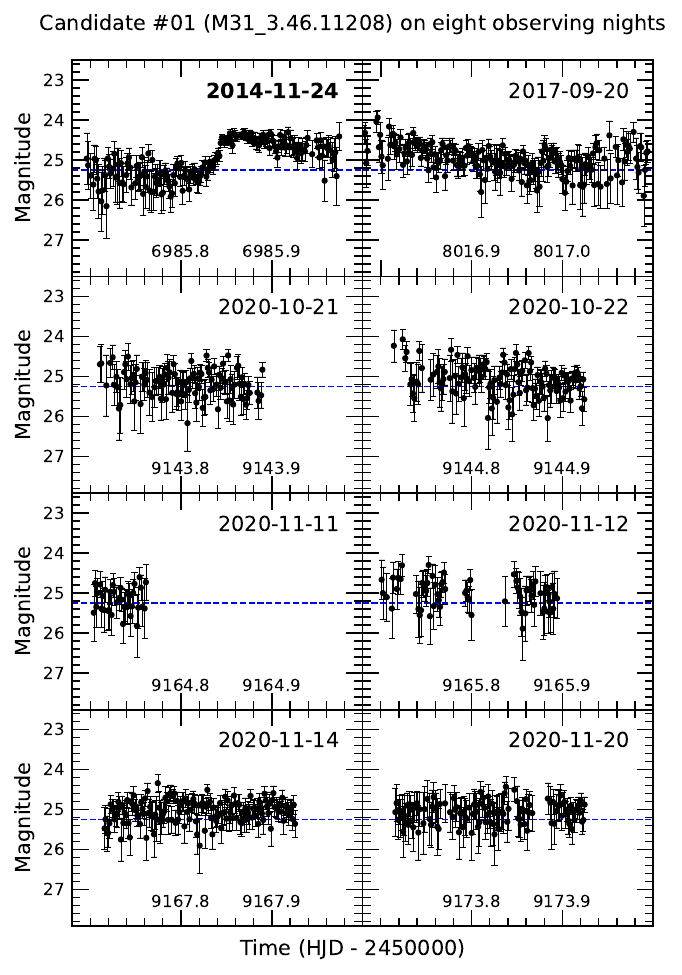}
\caption{Full Subaru HSC light curve of the candidate microlensing event \#01 found by \citet{sugiyama2026}. Each panel shows data from a separate night.}
\label{fig:lens11}
\end{figure}

\begin{figure}[h]
\includegraphics[width=\textwidth]{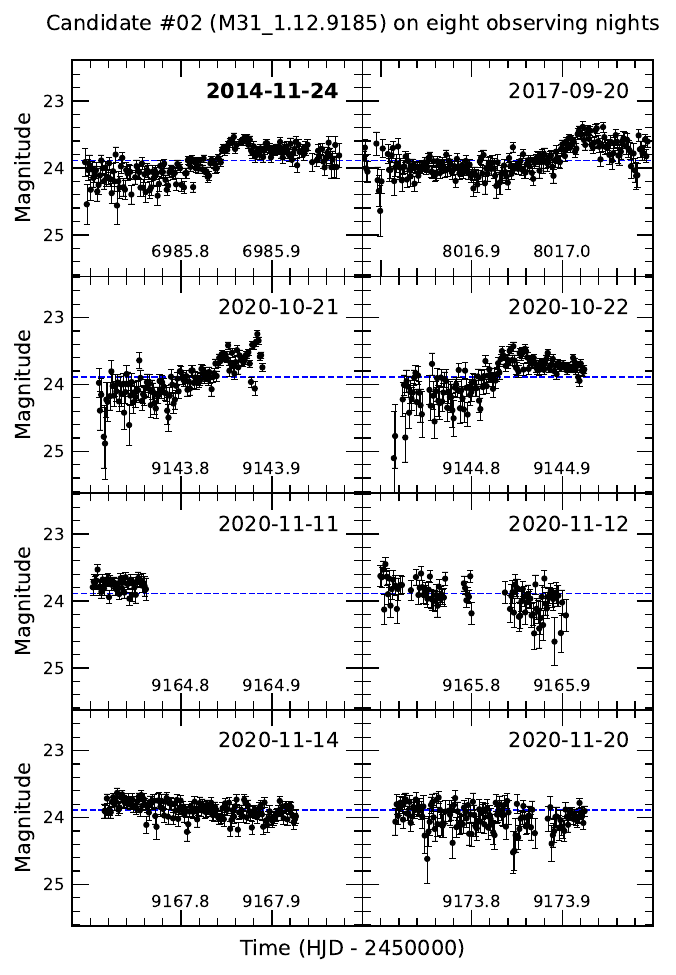}
\caption{Full Subaru HSC light curve of the candidate microlensing event \#02 found by \citet{sugiyama2026}. Each panel shows data from a separate night.}
\label{fig:lens12}
\end{figure}

\begin{figure}[h]
\includegraphics[width=\textwidth]{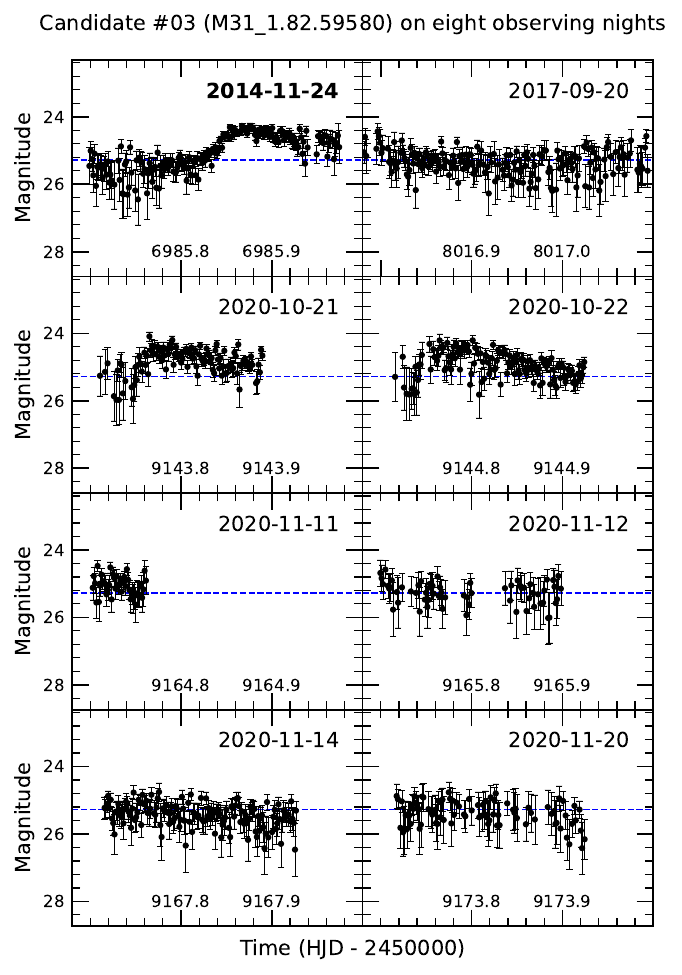}
\caption{Full Subaru HSC light curve of the candidate microlensing event \#03 found by \citet{sugiyama2026}. Each panel shows data from a separate night.}
\label{fig:lens13}
\end{figure}

\begin{figure}[h]
\includegraphics[width=\textwidth]{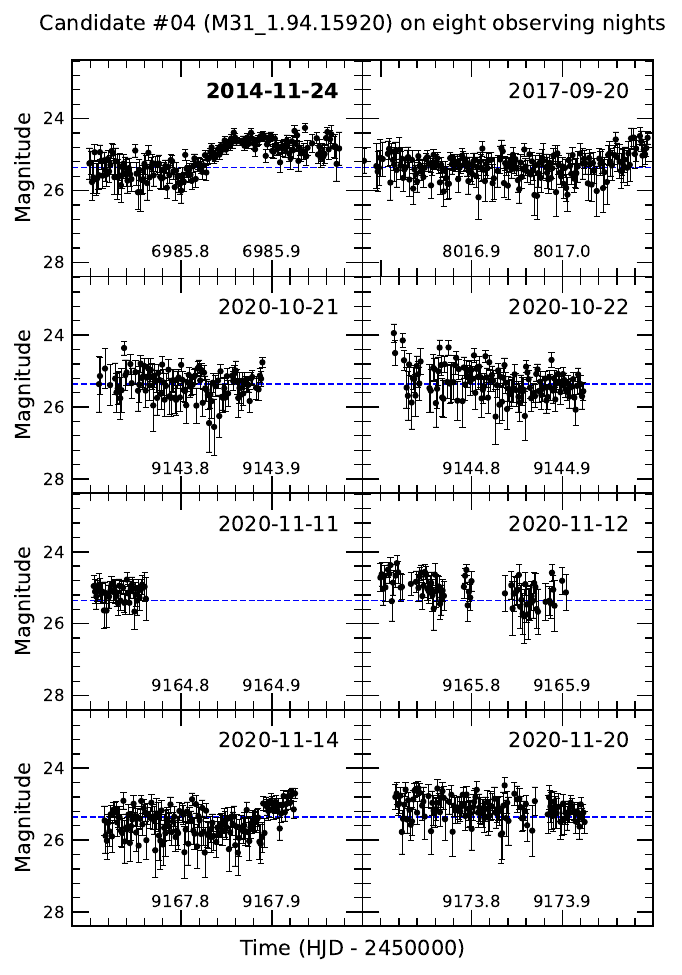}
\caption{Full Subaru HSC light curve of the candidate microlensing event \#04 found by \citet{sugiyama2026}. Each panel shows data from a separate night.}
\label{fig:lens14}
\end{figure}

\begin{figure}[h]
\includegraphics[width=\textwidth]{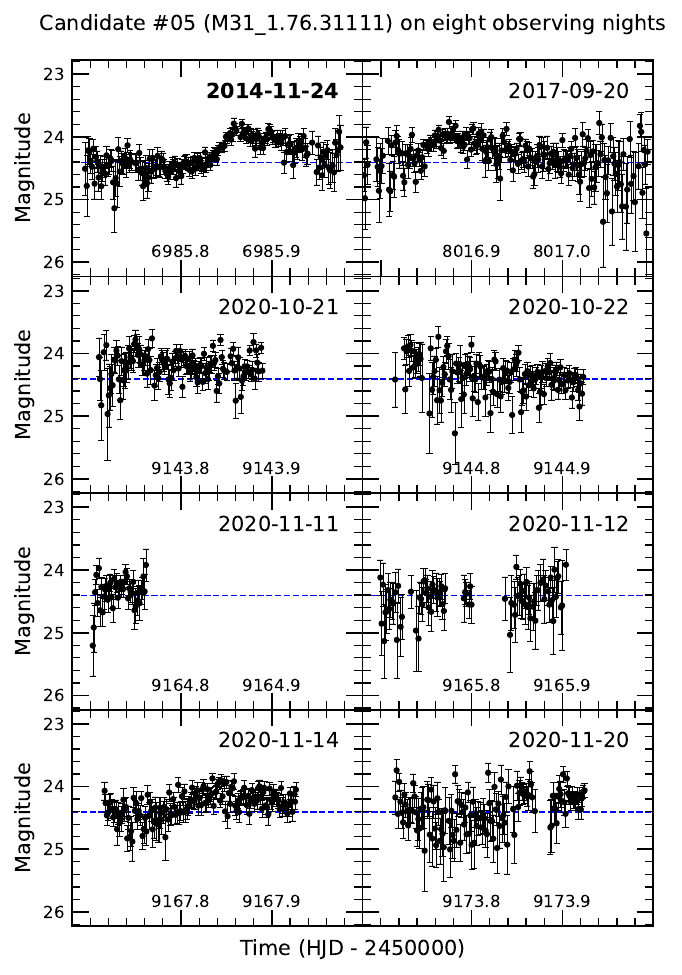}
\caption{Full Subaru HSC light curve of the candidate microlensing event \#05 found by \citet{sugiyama2026}. Each panel shows data from a separate night.}
\label{fig:lens15}
\end{figure}

\begin{figure}[h]
\includegraphics[width=\textwidth]{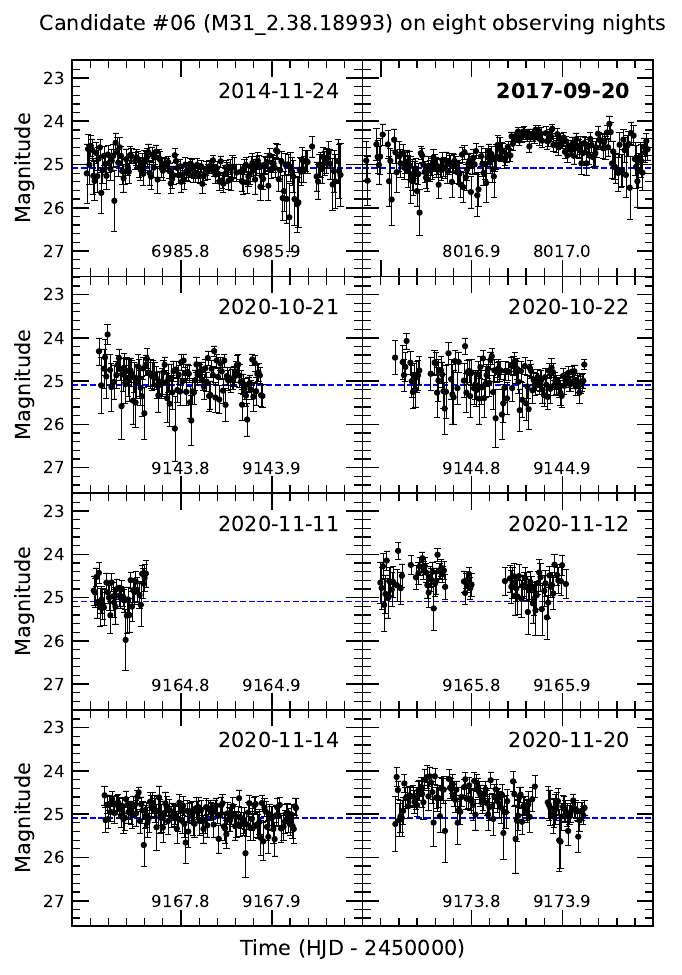}
\caption{Full Subaru HSC light curve of the candidate microlensing event \#06 found by \citet{sugiyama2026}. Each panel shows data from a separate night.}
\label{fig:lens16}
\end{figure}

\begin{figure}[h]
\includegraphics[width=\textwidth]{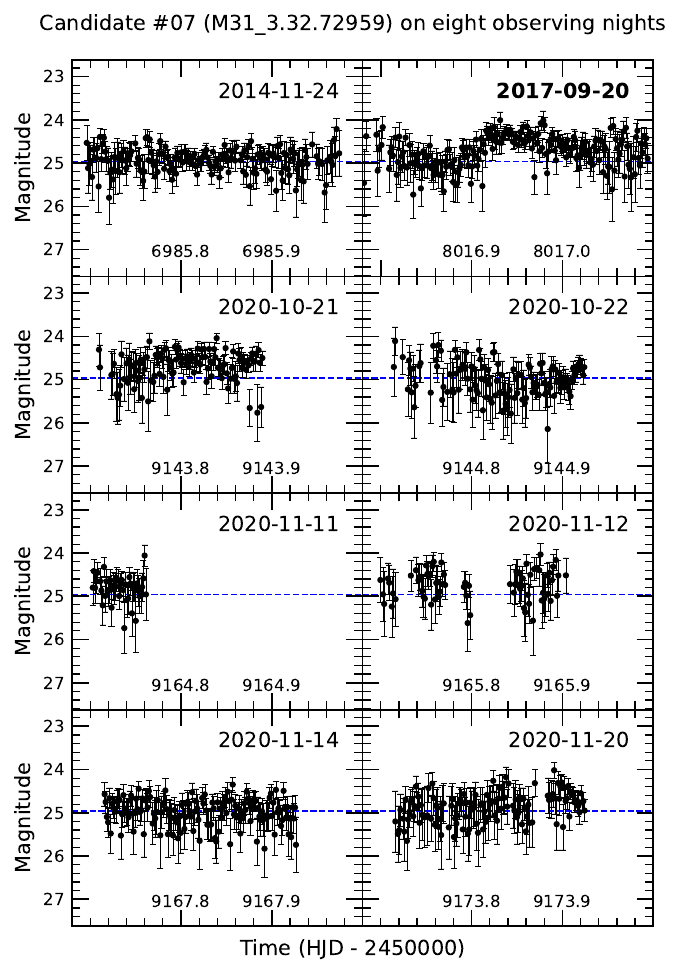}
\caption{Full Subaru HSC light curve of the candidate microlensing event \#07 found by \citet{sugiyama2026}. Each panel shows data from a separate night.}
\label{fig:lens17}
\end{figure}

\begin{figure}[h]
\includegraphics[width=\textwidth]{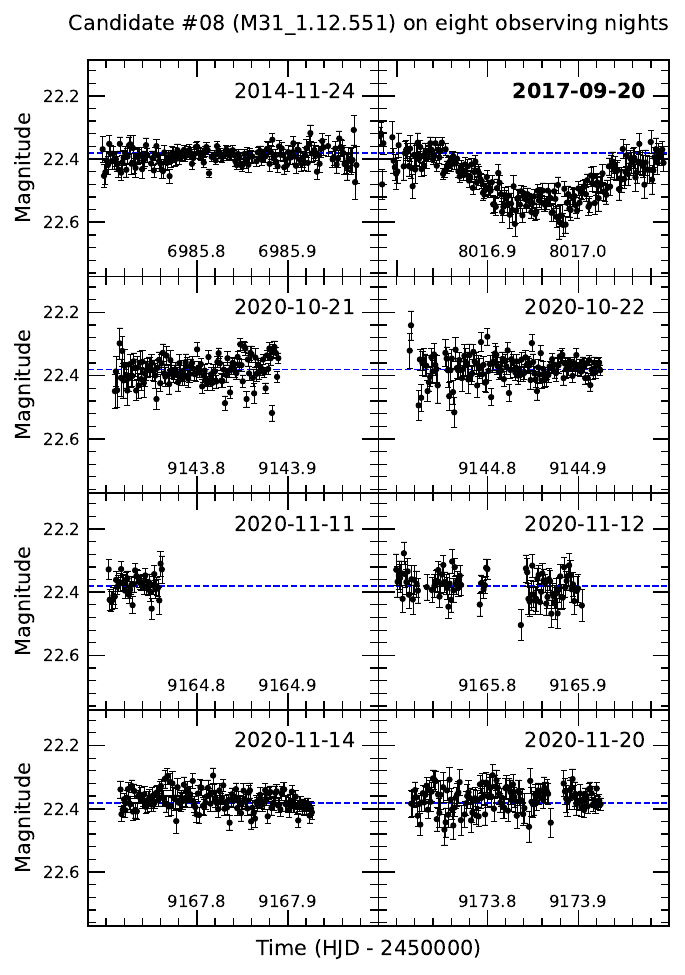}
\caption{Full Subaru HSC light curve of the candidate microlensing event \#08 found by \citet{sugiyama2026}. Each panel shows data from a separate night.}
\label{fig:lens18}
\end{figure}

\begin{figure}[h]
\includegraphics[width=\textwidth]{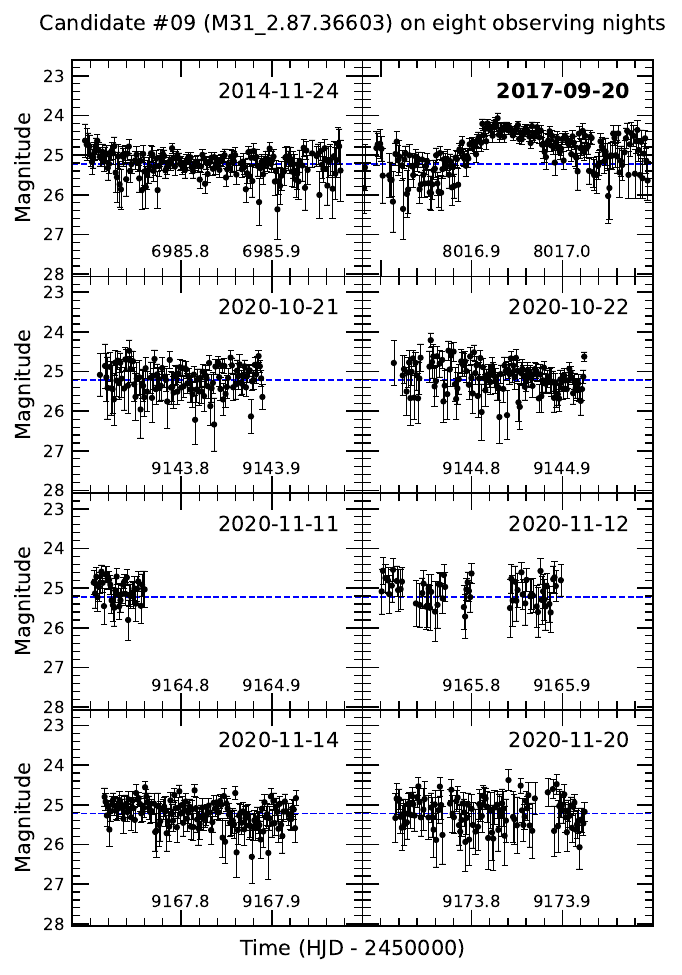}
\caption{Full Subaru HSC light curve of the candidate microlensing event \#09 found by \citet{sugiyama2026}. Each panel shows data from a separate night.}
\label{fig:lens19}
\end{figure}

\begin{figure}[h]
\includegraphics[width=\textwidth]{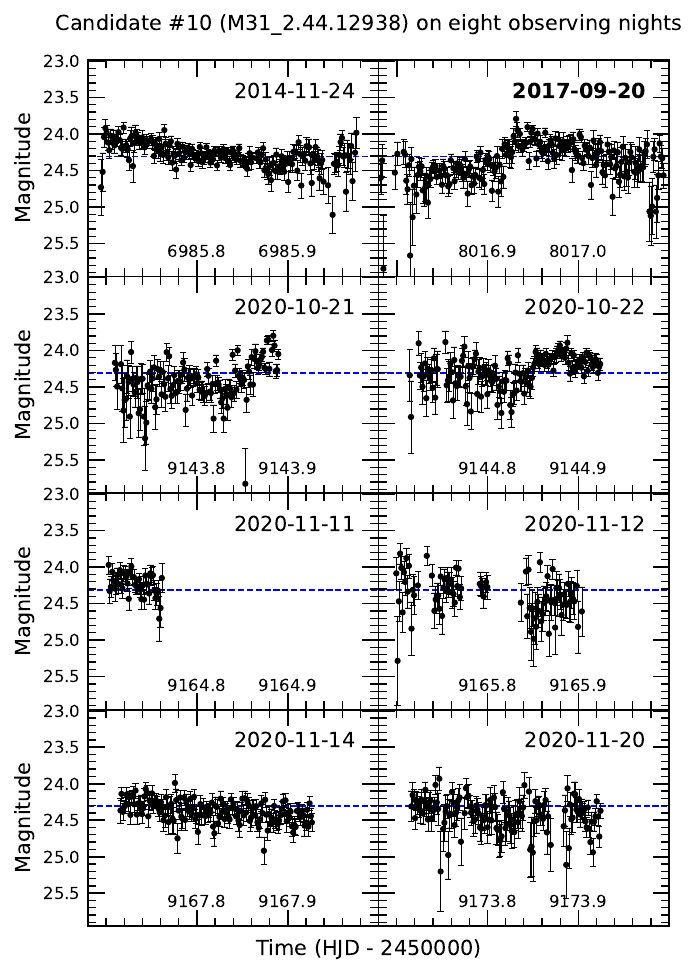}
\caption{Full Subaru HSC light curve of the candidate microlensing event \#10 found by \citet{sugiyama2026}. Each panel shows data from a separate night.}
\label{fig:lens20}
\end{figure}

\begin{figure}[h]
\includegraphics[width=\textwidth]{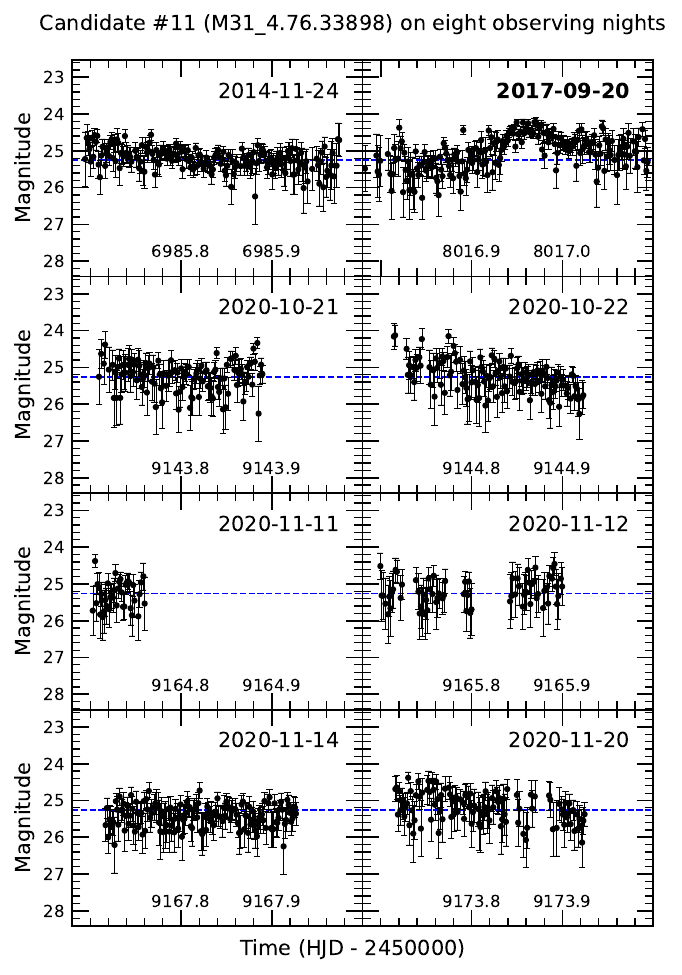}
\caption{Full Subaru HSC light curve of the candidate microlensing event \#11 found by \citet{sugiyama2026}.  Each panel shows data from a separate night.}
\label{fig:lens21}
\end{figure}

\begin{figure}[h]
\includegraphics[width=\textwidth]{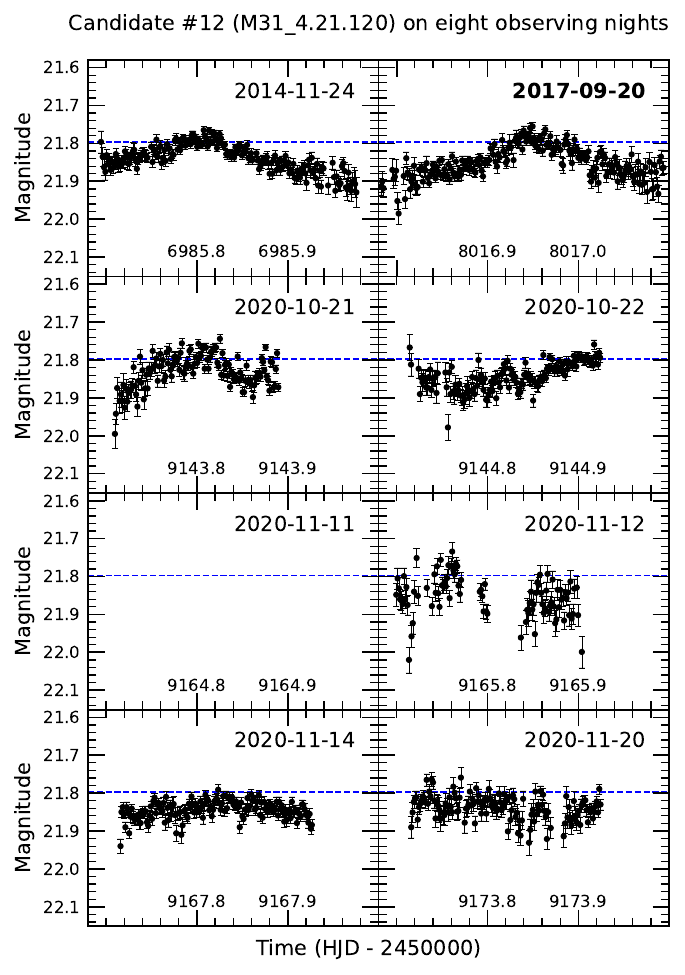}
\caption{Full Subaru HSC light curve of the candidate microlensing event \#12 found by \citet{sugiyama2026}. Each panel shows data from a separate night.}
\label{fig:lens22}
\end{figure}

Our full Subaru HSC light curves for all candidates are shown in Figures~\ref{fig:lens11}--\ref{fig:lens22}. Each panel shows data from a single night of observations. The night during which the event was originally detected by the \citet{sugiyama2026} pipeline is indicated in bold typeface. The horizontal blue dashed lines mark the brightness of the star measured in the reference image (epoch: ${\rm HJD'} = 6985.82$).

The light curves presented in Figures~\ref{fig:lens11}--\ref{fig:lens22} have two striking properties. First, many of the purported microlensing events have highly asymmetric light curves: the rise to the peak is much faster than the decline to the baseline brightness, which stands in stark contrast to the symmetric profiles expected from single-lens microlensing events. In addition, the variability of the candidates can also be clearly detected during other nights of Subaru monitoring.

Except for candidates \#8 and \#12, the light curves of all candidates are consistent with those of RR Lyrae stars. RR Lyrae variables are old, low-mass, radially pulsating stars with periods ranging from 0.2 to 1 day. Those pulsating in the fundamental mode have asymmetric light curves with a steep rising branch and a slow decline after maximum brightness. We refer readers to the OGLE Atlas of Variable Star Light Curves\footnote{https://ogle.astrouw.edu.pl/atlas/} for example light curves of such stars \citep[e.g.,][]{soszynski2014,soszynski2016}. RR Lyrae stars have typical $r$-band absolute magnitudes in the range $0\!-\!1.5$ mag \citep[e.g.,][]{narloch2024}. At the distance of M31 (distance modulus of $24.407 \pm 0.032$ mag; \citealt{li2021}), their expected apparent magnitudes fall between $24.5$ and $26$ mag, which matches the magnitudes of the objects identified by \citet{sugiyama2026}.

\begin{figure}[h]
\includegraphics[width=\textwidth]{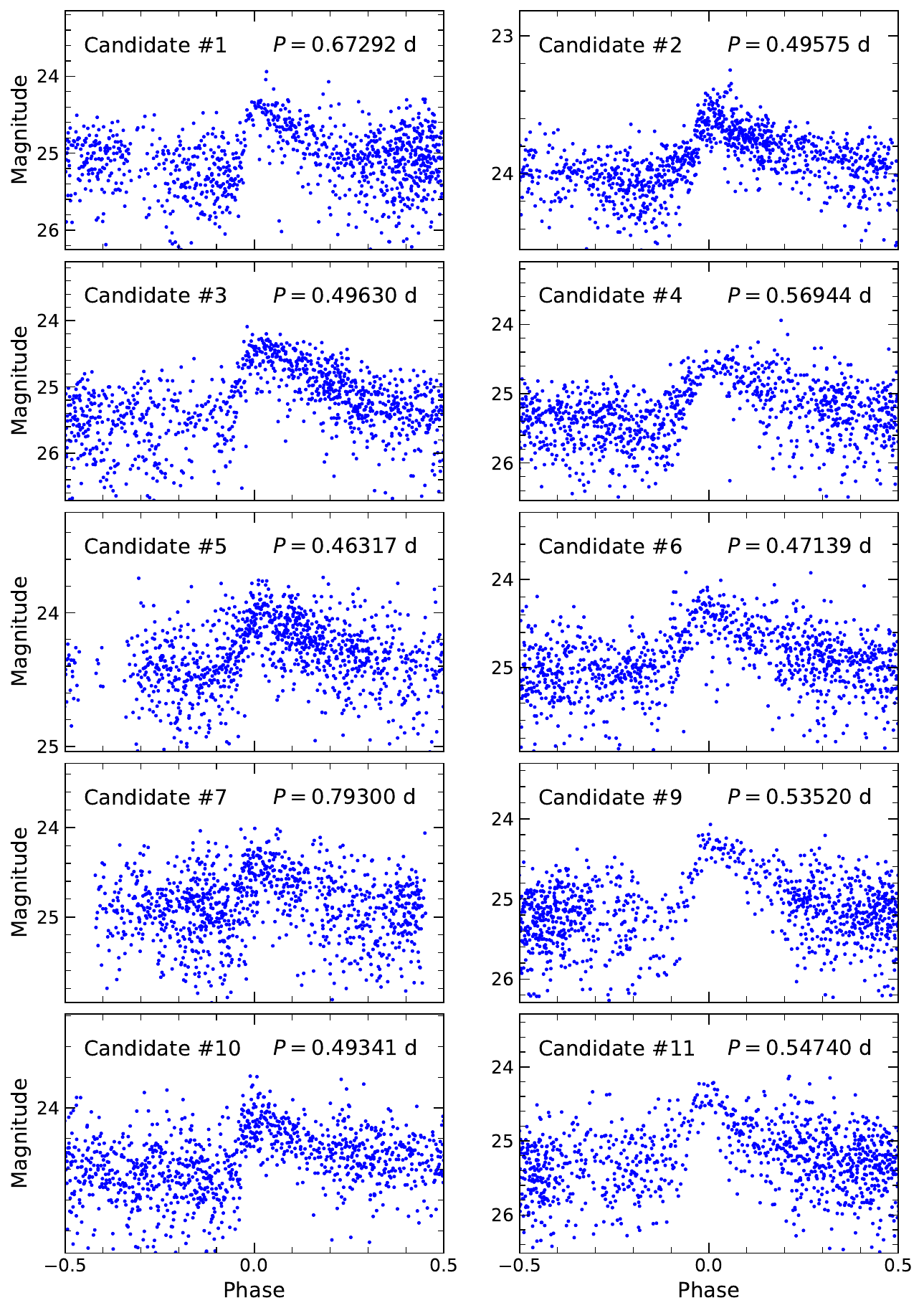}
\caption{Phase-folded light curves for ten candidates from \citet{sugiyama2026}, reclassified here as RR Lyrae stars.}
\label{fig:phased_all}
\end{figure}

As with the \citet{niikura2019} event candidate (Section~\ref{sec:niikura}), the sparse light curve coverage leads to period aliasing, making it difficult to uniquely determine the pulsation period. We used the analysis of variance algorithm \citep{aov1989} to search for possible periods in the data. We selected the period that yielded the highest S/N and reported it in Table~\ref{tab:events}. However, all reported periods remain tentative and require additional observations for confirmation. The phase-folded light curves for these ten candidates are shown in Figure~\ref{fig:phased_all}.

In addition to the ten RR Lyrae stars, candidate \#8 is a mundane eclipsing binary, while candidate \#12 is clearly a variable star. While the sparse light curve prevents an unambiguous classification, it may be a Cepheid. Our classifications for all objects are summarized in the last column of Table~\ref{tab:events}.

Given the clear variable star nature of the objects identified by \citet{sugiyama2026}, their original classification as microlensing event candidates warrants closer scrutiny. The ten clear RR Lyrae stars mistaken for microlensing candidates show distinctly asymmetric light curves, even in the photometric reductions shown in Figure~10 of \citet{sugiyama2026}. Although their event detection pipeline included some measures of the asymmetry in the light curves, they were apparently insufficient to distinguish the ``saw-tooth'' shape of a pulsating variable from a symmetric profile expected for a microlensing event. 

It is even more puzzling that stars exhibiting variability on multiple nights of Subaru monitoring were classified as microlensing candidates. \citet{sugiyama2026} stated that ``To ensure that the events are not due to repeating variables, we performed the forced photometry at the position of selected events on different observation dates. We examined both the light curves and the images, and removed the events showing suspicious variations in the light curve across different dates. However, note that we did not remove the events if the apparent variation on a different observation date is attributable to poor image subtraction.'' This final, subjective manual test permitted the inclusion of several false positives in their final sample. We have verified that nearby stars do not show similar variability at the same time as the candidates, indicating that these variations cannot be attributed to ``poor image subtraction.''

Finally, identifying ``microlensing candidates'' as RR Lyrae stars provides a natural explanation for their otherwise anomalous spatial distribution. Microlensing events are expected to be concentrated within the M31 disk, while RR Lyrae stars are found in large numbers in galactic halos. Because these stars are relatively faint ($24.5\!-\!26$~mag) and near the Subaru detection limit, they were difficult to identify during the 2020 observing run when the quality of the images was relatively poor. This explains why all \citet{sugiyama2026} ``events'' occurred during just two nights. It is less clear why their peaks are so clustered in time. We suspect that this may be an artifact of the event selection cuts introduced by \citet{sugiyama2026}.

The light curves of all event candidates are publicly available {\it via}
\begin{center}
\textit{https://www.astrouw.edu.pl/ogle/m31/}
\end{center}
or
\begin{center}
\textit{https://zenodo.org/records/19344727}
\end{center}

\section{Conclusions}

Our reanalysis of the Subaru HSC observations of M31 indicates that all microlensing event candidates identified by \citet{sugiyama2026} are, in fact, variable stars. This data set contains no compelling short-timescale microlensing events. Therefore, the Subaru observations do not provide evidence for a significant population of planetary-mass PBHs as dark matter components. Instead, the results presented by \citet{sugiyama2026} should be interpreted solely as upper limits.

\begin{figure}[h]
\includegraphics[width=\textwidth]{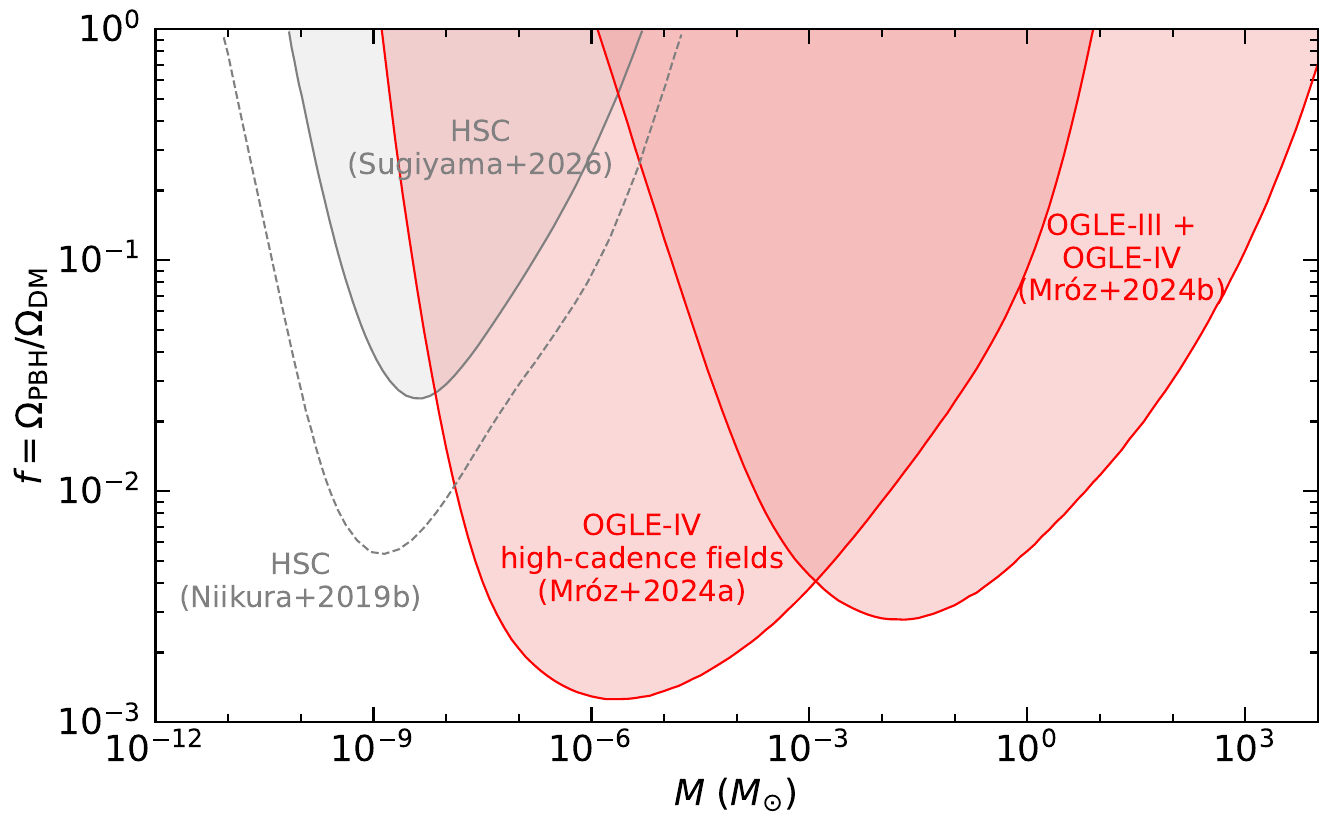}
\caption{The most stringent 95\% upper limits from the gravitational microlensing studies on the fraction of dark matter in the form of PBHs and other compact objects ($f=\Omega_{\rm PBH}/\Omega_{\rm DM}$). The shaded red regions mark limits from \citet[][\mbox{OGLE-IV} high-cadence fields]{mroz2024d} and \citet[][\mbox{OGLE-III} + \mbox{OGLE-IV}]{mroz2024b}. The limits found by \citet{sugiyama2026} and \citet{niikura2019} are marked with solid and dashed gray lines, respectively.}
\label{fig:bounds}
\end{figure}

These conclusions align well with the findings from searches for short-timescale microlensing events toward the Magellanic Clouds. Based on high-cadence OGLE observations, \citet{mroz2024d} provided stringent upper limits on the abundance of planetary-mass PBHs in dark matter. Specifically, OGLE observations indicate that PBHs with masses between $10^{-7}$ and $10^{-6}$\,M$_{\odot}$ constitute no more than 0.2\% of the dark matter in the Milky Way (Figure~\ref{fig:bounds}).

It is worth noting that while short-timescale events have not been detected in the directions of the Magellanic Clouds or M31, modern microlensing surveys have found evidence for a substantial population of such events toward the Milky Way bulge \citep{mroz2017,gould2022,sumi2023}. The most likely explanation for the bulge events is a population of free-floating or wide-orbit planets, although some studies \citep{niikura2019b} have suggested a population of planetary-mass PBHs. If such PBHs existed, their distribution in the Milky Way would follow that of the Galactic halo, and short-timescale microlensing events would be seen across the entire sky. In contrast, planets (whether free-floating or bound) should follow the distribution of stars.

Another notable aspect of the work of \citet{sugiyama2026} is that the inferred upper limits on the PBH abundance (calculated under the assumption that all candidate events are false positives, which---as we have demonstrated---is the case) are weaker than those in the earlier study by \citet{niikura2019}, see Figure~\ref{fig:bounds}. This is surprising given that the latter work used only a subset of the data analyzed by \citet{sugiyama2026}. For a given PBH mass, the upper limits on the PBH abundance in dark matter are inversely proportional to the number of expected microlensing events in the experiment, which is the product of the event detection efficiency, the expected event rate, the duration of the experiment, and the number of source stars. We agree that the most likely explanation for the difference between the \citet{niikura2019} and \citet{sugiyama2026} studies is the inclusion of finite-source effects in event detection simulations by the latter work. (Finite-source effects occur when the angular radius of the lensed star becomes larger than the angular Einstein radius, and may substantially reduce the number of observable events from planetary-mass lenses.)

In particular, the upper limits obtained by \citet{sugiyama2026} are stronger than those inferred by \citet{mroz2024d} for PBH masses lower than $10^{-8}$\,M$_{\odot}$, as shown in Figure~\ref{fig:bounds}. Because M31 is located much farther than the Magellanic Clouds, the potential source stars have smaller angular diameters, which reduces the impact of finite-source effects. Therefore, the high-cadence observations of M31 presented by \citet{sugiyama2026} still hold significant scientific value. It remains to be seen how robust these limits are, given that the selection criteria used to calculate detection efficiency permitted a substantial number of false positives and included some manual, subjective cuts. As we have already adapted the OGLE photometric pipeline for the reduction of the Subaru data, we plan to reanalyze the entire Subaru M31 dataset in the near future.

Our findings also serve as a cautionary tale for future microlensing surveys. They indicate that high cadence of observations alone may be insufficient to remove all possible false positives, and a long timespan of observations is necessary to eliminate the variable star contamination, especially when probing populations of faint ($\approx 25\!-\!26$\,mag) stars with large telescopes. The ongoing (DECam, Rubin) and planned (Roman) microlensing experiments may face similar challenges regarding false-positive contamination.

\section*{Acknowledgments}

We thank members of the OGLE team for discussions about M31 microlensing event candidates. This research was funded in part by National Science Centre, Poland, grant SONATA 2023/51/D/ST9/00187 awarded to P.M. A.U. acknowledges support from the grant OPUS-28 2024/55/B/ST9/00447 of the Polish National Science Centre.
This research is based on data collected at the Subaru Telescope, which is operated by the National Astronomical Observatory of Japan. We are honored and grateful for the opportunity of observing the Universe from Maunakea, which has the cultural, historical, and natural significance in Hawaii. The data were obtained from the SMOKA, which is operated by the Astronomy Data Center, National Astronomical Observatory of Japan. For the purpose of Open Access, the authors have applied a CC-BY public copyright license to any Author Accepted Manuscript (AAM) version arising from this submission.

\bibliographystyle{acta}
\bibliography{pap}

\end{document}